\documentstyle[twoside,fleqn,espcrc2]{article}
\input{psfig}
\def\bold#1{\setbox0=\hbox{$#1$}%
     \kern-.025em\copy0\kern-\wd0
     \kern.05em\copy0\kern-\wd0
     \kern-.025em\raise.0433em\box0 }
\def\slash#1{\setbox0=\hbox{$#1$}#1\hskip-\wd0\dimen0=5pt\advance
       \dimen0 by-\ht0\advance\dimen0 by\dp0\lower0.5\dimen0\hbox
         to\wd0{\hss\sl/\/\hss}}

\newcommand{\be}{\begin{equation}}
\newcommand{\ee}{\end{equation}}
\newcommand{\bea}{\begin{eqnarray}}
\newcommand{\eea}{\end{eqnarray}}

\newcommand{\AmS}{{\protect\the\textfont2
  A\kern-.1667em\lower.5ex\hbox{M}\kern-.125emS}}

\title{Universal Isgur-Wise form factors from QCD sum rules in HQET}

\author{P. Colangelo$^a$\thanks{speaker at the Conference},
F. De Fazio$^a$ and N. Paver$^b$\\
\vskip 0.2cm
$^a$ Istituto Nazionale di Fisica Nucleare - Sezione di Bari, 
via Amendola n.173, 70126 Bari, Italy\\ 
\vskip 0.2cm
$^b$ Dipartimento di Fisica Teorica dell'Universit\'a di Trieste and
Istituto Nazionale di Fisica Nucleare - Sezione di Trieste, 
Strada Costiera 11, 34014 Trieste, Italy\\}

\begin{document}

\begin{abstract}
We review the role of the universal Isgur-Wise functions 
parameterizing the $B$ meson 
semileptonic  matrix elements to excited charm states 
in the infinite heavy quark mass limit. We also discuss the 
determination of such form factors by QCD sum rules in the framework of 
the Heavy Quark Effective Theory.

\end{abstract}

\maketitle

\section{UNIVERSAL FORM FACTORS}

The heavy quark flavor and spin 
symmetry of QCD, exactly valid  in the infinite $m_Q$ limit, 
has important consequences on the spectrum, the static properties and both
the strong and weak decay matrix elements of the hadrons
containing a single heavy quark $Q$ \cite{hqet_1,hqet_2,reviews}. 
Notably, in such heavy quark effective theory (HQET),
for $m_Q \to \infty$ the states can be classified in degenerate doublets
by the total angular momentum $\vec J$ and by
the angular momentum of the light degrees of freedom
(quarks and gluons) $\vec s_\ell= \vec J - \vec s_Q$, and the 
semileptonic matrix elements between members of the doublets identified by
$s_\ell$ and $s_{\ell^\prime}$ can be expressed in terms 
of ``universal'' form factors which are functions of
$y=v \cdot v^\prime$ (with $v$ and $v^\prime$ the initial and final hadron  
four-velocities). 
This is a well known result for
$B \to D^{(*)} \ell \bar\nu $:
the pseudoscalar  $B,D$ and vector
$B^*,D^*$  mesons have $s_\ell^P={1\over2}^-$,
and the semileptonic  matrix elements can be written
in terms of a single  
universal form factor, the Isgur-Wise function $\xi(y)$.
The heavy flavour symmetry requires that such form factor is 
normalized to unity at the zero recoil point $y=1$.

In the case of transitions between members belonging to different
HQET  multiplets, additional form factors are introduced.
An important example involves
the four meson states corresponding
to the orbital angular momentum $L=1$ 
($P$ waves in the constituent quark model), which can be classified 
in the doublets: 
$J^P=(0^+_{1/2},1^+_{1/2})$ and
$J^P=(1^+_{3/2},2^+_{3/2})$, according to the values 
$s_\ell^P={1\over 2}^+$ and $s_\ell^P={3 \over 2}^+$, respectively.
Denoting the corresponding charm mesons  by  
$(D_0,D_1^*)$ and $(D_1,D_2^*)$,
the weak matrix elements  
$<(D_0, D_1^*)| V-A | B>$ and $<(D_1, D_2^*)| V-A | B>$ can be written
in terms of two independent functions, 
$\tau_{1/2}$ and $\tau_{3/2}$, respectively
\cite{isgur91}, that
allow the complete description
of the physical processes at the leading order in $1/m_Q$.
However, contrary to the case of  the Isgur-Wise function
$\xi$, one cannot invoke symmetry 
arguments to predict the 
normalization  at $y=1$ of 
$\tau_{1/2}(y)$ and $\tau_{3/2}(y)$.

From the phenomenological point of view, the $B \to D^{**}$ 
semileptonic transitions ($D^{**}$ being the generic $L=1$ charmed state)
are interesting, since in principle these decay modes may  
account for a sizeable fraction of the inclusive semileptonic $B$-decay
rate. In any case they represent a well-defined set of corrections to the 
theoretical prediction that, in the limit 
$m_Q\to\infty$ and under the condition  $(m_b-m_c)/(m_b+m_c)\to 0$,
the total semileptonic $B\to X_c$ decay rate should be saturated by 
the $B\to D$ and $B\to D^*$ modes \cite{hqet_2}. 
Another point of interest,  
the values  of the $B\to D^{**}$ form factors at $y=1$  
provide a lower bound to the slope $\rho^2$ of the function $\xi$ \cite{bj90}:
\be
\rho^2 \ge {1\over 4}+ |\tau_{1/2} (1)|^2 + 2  |\tau_{3/2} (1)|^2 \;\;\;.
\ee
They also appear in the ``dipole'' sum rule  \cite{vol92}
\be
{\bar \Lambda \over 2} = 
( \bar \Lambda^+ -\bar \Lambda) |\tau_{1/2}(1)|^2 +
2 ( \bar \Lambda^{T} -\bar \Lambda)
|\tau_{3/2}(1)|^2+ \dots 
\ee 
where  
$\bar \Lambda=M_B - m_b$, 
$\bar \Lambda^+=M_{D_0} - m_c$,
$\bar \Lambda^T=M_{D_1} - m_c$ are heavy meson  ``binding energies''.

The investigation of the
semileptonic $B$ transitions to excited charm states
is an important issue for the theoretical analysis of the $D^{**}$ 
production  in nonleptonic $B$ decays \cite{nonlepb},
as well as for CP physics at $B$ factories \cite{babar}. As for
$\Lambda_{QCD}/m_Q$ effects in such processes, they are discussed in
\cite{ligeti97}.     

The charmed $2^+_{3/2}$ state, 
$D_2^*(2460)$, has been observed with 
$m_{D_2^*}=2458.9\pm 2.0$ MeV, $\Gamma_{D_2^*}=23\pm 5$ MeV and
$m_{D_2^*}=2459\pm 4$ MeV, $\Gamma_{D_2^*}=25^{+8}_{-7}$ MeV for the neutral 
and charged states, respectively. 
The HQET state $1^+_{3/2}$ can be identified with $D_1(2420)$, with
$m_{D_1}=2422.2\pm 1.8$ MeV and 
$\Gamma_{D_1}=18.9^{+4.6}_{-3.5}$  MeV  
(a $1^+_{1/2}$ component is contained in this physical state,
due to the mixing allowed for the finite value of the charm quark mass). 
There is also some experimental evidence of beauty
$s_\ell^P={3\over 2}^+$ states \cite{pdg}.

Both the states $2^+_{3/2}$ and $1^+_{3/2}$ decay to hadrons by
$d-$wave suppressed transitions.
On the other hand, the $s_\ell^P={1\over 2}^+$ doublet ($D_0, D^*_1$) 
has not been observed yet.
The strong decays of such states occur through $s$-wave transitions,
with expected larger
widths than in the case of the 
doublet ${3\over 2}^+$. Theoretical analyses of 
the coupling constants governing the two-body hadronic transitions 
can be carried out by QCD sum rules, in analogy with the determinations of 
the $B^* B \pi$, $D^* D \pi$ couplings \cite{narison}, obtaining
$\Gamma(D_0^0 \to D^+ \pi^-) \simeq 180$ MeV and
$\Gamma(D_1^{*0} \to D^{*+} \pi^-) \simeq 165$ MeV, 
and the mixing 
angle $\alpha$ between $D^*_1$ and  $D_1$  
$\alpha \simeq 16^0$  \cite{colangelo95}.

\section{FORM FACTOR $\tau_{1/2}$ FROM QCD SUM RULES IN HQET}

The matrix elements of the semileptonic 
$B \to D_0 \ell \bar \nu $ and $B \to D_1^* \ell \bar \nu$  
transitions can be parameterized in terms of six form factors: 
\bea
{ <D_0(v')|{\bar c}\gamma_\mu \gamma_5 b|B(v)> 
\over {\sqrt {m_B m_{D_0}}}  } &=& \nonumber\\
g_+(y) (v+v^\prime)_\mu +
 g_-(y) (v-v^\prime)_\mu \;\;\;&,& \nonumber 
\eea
\bea
{<D^*_1(v',\epsilon)|{\bar c}\gamma_\mu(1-\gamma_5)b|B(v)>
\over {\sqrt {m_B m_{D^*_1}}} } &=& \nonumber \\
g_{V_1}(y) \epsilon^*_\mu +\epsilon^* \cdot v \; [g_{V_2}(y) v_\mu+
g_{V_3}(y) v^\prime_\mu]&-& \nonumber \\
i \; g_A(y)
\epsilon_{\mu \alpha\beta \gamma} \epsilon^{*\alpha} v^\beta v^{\prime \gamma}
\;\;\; &.&  \label{full}
\eea
The relation of  $g_i(y)$ in 
(\ref{full}) to  $\tau_{1/2}(y)$
\cite{isgur91}, in the limit 
$m_Q\to\infty$,  involves short-distance coefficients
which depend on the heavy quark masses $m_b,m_c$,
on $y$ and on a mass-scale $\mu$, and connect the QCD vector and axial vector 
currents to the HQET currents. At the next-to-leading 
logarithmic approximation in  $\alpha_s$ 
such relations  are: 
\bea
g_+ + g_- &=& -2 \;\Big( C_1^5+(y-1) C_2^5\Big) 
\; \tau_{1/2} \nonumber \\
g_+ - g_- &=& 2\;\Big( C_1^5-(y-1)C_3^5\Big)
 \; \tau_{1/2}\nonumber \\
g_{V_1}&=&2(y-1) \;C_1 \; \tau_{1/2} \nonumber \\
g_{V_2}&=&-2\; C_2\; \tau_{1/2} \nonumber \\
g_{V_3}&=&-2\;\Big(C_1+C_3\Big) \;\tau_{1/2} \nonumber \\
g_A&=&-2 \; C_1^5 \; \tau_{1/2}\;. 
\label{rel_formf}
\eea
The Wilson coefficients $C_i$, depending on a scale $\mu$, are reported, 
e.g., in \cite{colangelo98}.
Since the form factor $\tau_{1/2}$ is defined by the 
matrix elements of weak currents in the effective theory,
it depends on $\mu$;
it is possible to remove the 
scale-dependence by 
compensating it by the  $\mu$-dependence of the Wilson coefficients, 
defining $\tau_{1/2}^{\rm ren}$.

The universal functions can be estimated by  
non-perturbative approaches; 
a genuinely field theoretical method is represented by  
 HQET QCD sum rules, which allow to
relate  hadronic observables to QCD parameters {\it via}
the Operator Product Expansion (OPE) of suitable 
Green functions. The expansion involves
$\alpha_s$ corrections in the coefficients of the OPE, as well 
as non-perturbative quark and gluon vacuum condensates. 

A critical aspect of the sum rule calculations in HQET
is represented by the size of non-leading terms,  such as
the $\alpha_s$ corrections. For example, 
the predictions for the leptonic constants 
of $\bar q Q$ pseudoscalar mesons
are affected by considerably 
large next-to-leading corrections in $\alpha_s$ 
\cite{reviews}. 
Conversely, for the Isgur-Wise function $\xi(y)$ 
the next-to-leading order $\alpha_s$ corrections 
turn out to be small and well under control \cite{neubert93}. 

In what follows we present a recent
determination of $\tau_{1/2}(y)$ in the frameworks of HQET QCD sum rules
at the next-to-leading order in 
the strong coupling $\alpha_s$ \cite{colangelo98}.
The method
is based on the three-point correlator 
\bea
\Pi_\mu(\omega, \omega^\prime, y)~&=&~
i^2~\int dx \; dz e^{i(k' x-k z)}  \nonumber \\
&<&0|T[ J_s^{v'}(x), {\tilde A}_\mu(0), J_5^{v}(z)^\dagger]|0> \;\;\;
 \label{threep}
\eea
with
${\tilde A}_\mu= \bar h_{Q^\prime}^{v'} \gamma_\mu \gamma_5 h_Q^v$ the 
$b \to c$ weak axial current in HQET, 
$J_s^{v'}=\bar q h_{Q^\prime}^{v\prime}$ and 
$J_5^{v}=\bar q i \gamma_5 h_Q^v$  
local  currents interpolating $D_0$ and $B$,
$h_Q^v$ being HQET quark fields. Here,
$\omega=2 v\cdot k$ and $\omega^\prime=2 v^\prime\cdot k^\prime$, and
 $k, k^\prime$ are 
residual momenta in the expansion of the heavy meson momenta:
$P=m_Q v+ k$, $P^\prime=m_{Q^\prime} v^\prime+ k^\prime$.
The residual momenta remain finite in the heavy quark limit.
 
Using the analyticity of $\Pi$ in 
$\omega$ and $\omega^\prime$ at fixed $y$, we can
represent (\ref{threep}) in terms of physical hadronic states,
giving poles at positive 
values of $\omega$ and $\omega^\prime$ and a continuum.
The lowest-lying contribution,  determined by the  
$B$, $D_0$ pole, introduces the form factor of interest:
\begin{equation}
\Pi_{pole}(\omega, \omega^\prime, y) = 
{-2 \tau_{1/2}(y, \mu)  F(\mu)  F^+(\mu)  \over 
(2 \bar \Lambda  - \omega - i \epsilon) 
(2 \bar \Lambda^+ - \omega^\prime - i \epsilon) } \;. \label{pole}
\end{equation}
$F(\mu)$, $F^+(\mu)$ 
are HQET couplings of the pseudoscalar and scalar interpolating 
currents to their respective  $0^-$ and $ 0^+$  bound states:
\begin{eqnarray} 
<0| J_5^v |B(v) > &=& F(\mu) \label{f} \\
<0| J_s^v |B_0(v) > &=& F^+(\mu) \label{fhat} \;\;\;.
\end{eqnarray}
Notice that $F(\mu)$ is related to the familiar
$B$-meson leptonic decay constant $f_B$. 
The parameters $\bar \Lambda$ and $\bar \Lambda^+$ identify the position 
of the poles in $\omega$ and $\omega^\prime$. 
As noted previously, 
$\mu$ independent quantities $F^+$ and $F$ can be obtained by  suitable 
perturbative factors.

The contribution of higher states to $\Pi$ in (\ref{threep}) is
taken into account by invoking the quark-hadron duality 
ansatz, and is modeled by a perturbative QCD continuum. 

The correlator (\ref{threep}), at large negative $\omega, \omega^\prime$,
can also be expressed in QCD 
by using the short-distance operator product expansion, in terms of 
perturbative and nonperturbative contributions:
\begin{equation}
\Pi = \Pi^{pert} + \Pi^{np}
\;\;\;. \label{qcd}
\end{equation}
In (\ref{qcd})
$\Pi^{np}$ represents the series of power corrections in  the "small"
variables ${1 \over \omega}$  and  ${1 \over \omega^\prime}$. These
corrections are determined by quark and gluon vacuum condensates, which
account for the features of 
the nonperturbative strong interactions.

The QCD sum rule 
for $\tau_{1/2}$ is obtained by imposing that the QCD 
and the hadronic representations  numerically
match in a suitable range of 
Euclidean values of $\omega$ and $\omega^\prime$. 

To "optimize" the sum rule,
a double Borel transform  in the variables 
$\omega$ and $\omega^\prime$ is applied. This allows to eliminate 
subtraction terms, improve
the convergence 
of the nonperturbative series and enhance
the role of the lowest-lying meson states.

\begin{figure}[htb]
\vskip -1.0cm
\psfig{figure=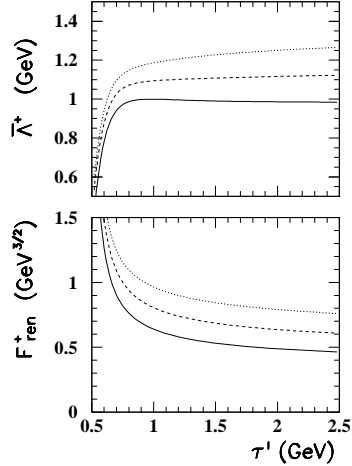,height=8.0 cm}
\vskip -1.5cm
\caption{$\bar \Lambda^+$ and  $F^+$ vs the Borel parameter $\tau^\prime$.}
\label{fig:fpiu}
\end{figure}

The same method can be applied to determine $F^+$, $\bar \Lambda^+$ 
(fig.\ref{fig:fpiu})
and $F$,  $\Lambda$, considering two-point correlators.
The result is
$\Lambda^+= 1.0 \pm 0.1 \;$ GeV,
$F^+=0.7 \pm 0.2$ GeV$^{3\over 2}$, and
$\Lambda= 0.5 \pm 0.1$ GeV, 
$F=0.45 \pm 0.05$ GeV$^{3\over 2}$. 
The difference $\Delta= \bar \Lambda^+ - \bar \Lambda$
corresponds to  
$\Delta=m_{\bar D_0}-m_{\bar D_0}$, with
$\bar D$ and  $\bar D_0$ the spin averaged states of the 
${1\over 2}^-$ and ${1\over 2}^+$ doublets. The central value
$\Delta=0.5 \; GeV$ enables the prediction $m_{\bar D_0}\simeq 2.45$ GeV,
with an uncertainty of about $0.15$ GeV.
Determinations of
$F^+$ at the order $\alpha_s=0$ by QCD sum rules  gave 
the result: $F^+=0.46 \pm 0.06$ GeV$^{3\over 2}$ \cite{colangelo92}
and $F^+=0.40 \pm 0.04$ GeV$^{3\over 2}$ and ${\bar \Lambda}^+=1.05 \pm 0.5$ 
GeV or ${\bar \Lambda}^+=0.90 \pm 0.10$ GeV \cite{dai} 
depending on the choice of the interpolating currents,
which shows the  significant size of the $\alpha_s$  
corrections in this case.

Determinations of $F^+$ by relativistic quark models give
$F^+ \simeq 0.6 - 0.7$ GeV$^{3/2}$ \cite{morenas97}.

\begin{figure}[htb]
%\vspace{9pt}
\vskip -1.0cm
\psfig{figure=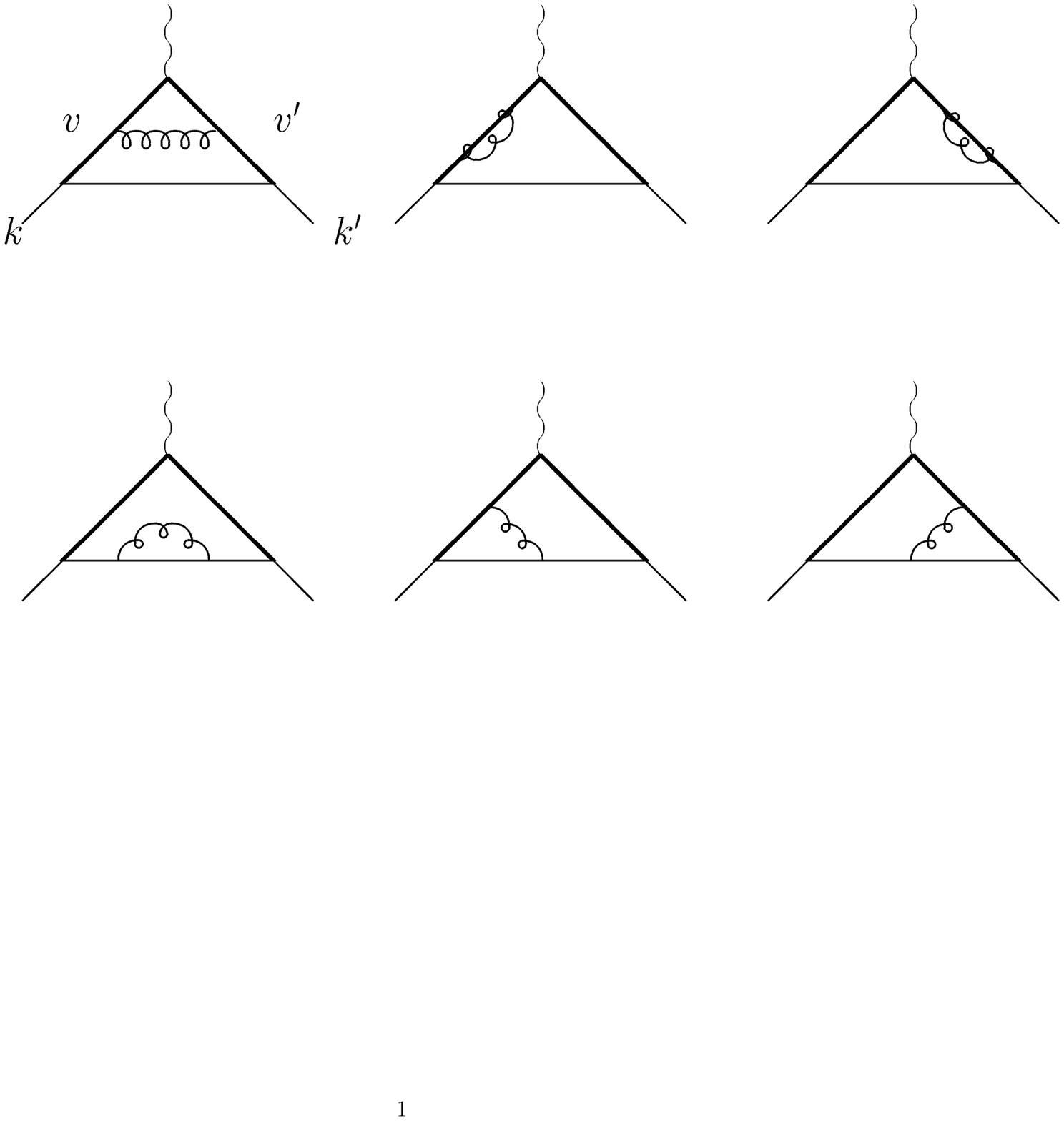,height=8.0 cm}
\vskip -4.5cm
\caption{Two-loop diagrams in $\Pi^{pert}$.}
\label{fig:triangles}
\end{figure}

The ${\cal O}(\alpha_s)$ corrections to the
perturbative part of the sum rule for 
$\tau_{1/2}$ are represented by the two-loop diagrams 
in fig.\ref{fig:triangles}. 
The result of their calculation, as well as of
${\cal O}(\alpha_s)$ corrections
to the  condensate terms, is reported in \cite{colangelo98}.

\begin{figure}[htb]
%\vspace{9pt}
\vskip -1.0cm
\psfig{figure=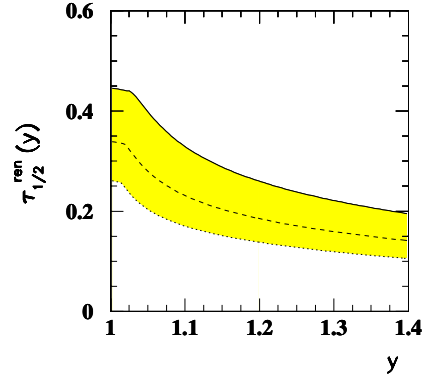,height=8.0 cm}
\vskip -2.5cm
\caption{Form factor $\tau_{1/2}(y)$  from QCD sum rules}
\label{fig:tau12}
\end{figure}

In the numerical analysis of the sum rule for $\tau_{1/2}$, 
the $\alpha_s$ contribution to $\Pi^{pert}$ turns out to be sizeable.
However, it is partially compensated by the analogous 
corrections in $F$ and $F^+$.
 This is a 
remarkable result, not expected a priori since the normalization of  
$\tau_{1/2}$ is not fixed by symmetry arguments.
The obtained  $\tau_{1/2}$ is depicted in fig.\ref{fig:tau12}, where
the shaded region essentially represents the 
theoretical uncertainty  of the calculation.
Using the expansion near $y=1$: 
\be
\tau_{1/2}^{ren}(y)=\tau_{1/2}(1) 
\Big(1-\rho^2_{1/2} (y-1)+c_{1/2} (y-1)^2\Big) 
\ee
we get
$\tau_{1/2}(1)=0.35\pm0.08$, 
$\rho^2_{1/2}=2.5\pm 1.0$ and 
$c_{1/2}=3\pm 3$.
Neglecting  $\alpha_s=0$ corrections QCD sum rules would  give  
$\tau_{1/2}(1) \simeq 0.25$ \cite{colangelo92,dai98}.
 
Other determinations of $\tau_{1/2}(y)$
have appeared in the literature, based on
various versions of the constituent quark model \cite{qmod}. 
The results range in a broad interval,
$\tau_{1/2}(1)=0.06-0.40$ and
$\rho^2_{1/2}=0.7-1.0$, and critically depend on the detailed
features of the models employed in the numerical calculation.

As for $\tau_{3/2}(y)$, a QCD sum rule analysis to the order 
$\alpha_s=0$  gives
$\tau_{3/2}(1)\simeq 0.28$ and $\rho^2_{3/2}\simeq 0.9$ \cite{colangelo92}.
Quark models give predictions in the range
$\tau_{3/2}(1)\simeq 0.31-0.66$ and $\rho^2_{3/2}\simeq 1.4-2.8$
\cite{qmod}. 

As far as the decays  $B \to (D_0, D_1^*) \ell \nu$ are concerned, using
$V_{cb}=3.9 \times 10^{-2}$ and 
$\tau(B)= 1.56$ ps, we predict
${\cal B}(B\to  D_0 \ell \bar \nu)=(5 \pm 3) \times 10^{-4}$ and
${\cal B}(B\to D^*_1 \ell \bar \nu)= (7 \pm 5) \times 10^{-4}$.
This represents  only a very small fraction of the semileptonic $B \to X_c$ 
decays . Although the branching ratios are small, one can hope that
the semileptonic $B$ transitions to the $s_\ell^P={1\over 2}^+$ charm doublet
will be identified at the $B$-factories. 
At present, the  $s_\ell^P={1\over 2}^+$ charm doublet
is difficult to be distinguished from the non-resonant background.

In conclusion, we stress that predictions derived within HQET
should always be supported by the computation of $1/m_Q$ as well as radiative 
corrections. The role of both depend on the various  situations.
In the case of $\tau_{1/2}$ we have obtained that $\alpha_s$ 
corrections are under control for $\tau_{1/2}$, while they  
considerably  affect the leptonic constant $F^+$.

\end{document}